\documentclass[11pt]{article}
\usepackage{latexsym}

\def\be{\begin{equation}}
\def\ee{\end{equation}}
\def\bea{\begin{eqnarray}}
\def\eea{\end{eqnarray}}

\def\case#1/#2{\textstyle\frac{#1}{#2}}

\begin{document}

\begin{center}

\Large{\bf SPACETIME OR SPACE AND THE PROBLEM OF TIME}\normalsize

\mbox{ }

\large{\bf E. Anderson\footnote{\uppercase{I} acknowledge 
PPARC studentship.}}\normalsize

\mbox{ }

\large{\em Astronomy Unit, Department of Mathematics, Queen Mary, London}\normalsize

\end{center}

\mbox{ }

\noindent In GR one may choose to use the spacetime action or its rearrangement\cite{ADM}\footnote{Here $h_{ij}$ is the induced metric of a 3-space with determinant $h$, Ricci 3-scalar $R$ 
and conjugate momentum $p^{ij}$ with trace $p$.  In moving between nearby 3-spaces, the 
lapse $\alpha$ is the time elapsed and the shift $\beta_i$ is the ammount of stretching of the 
3-space coordinate grid.}  
$$
\mbox{\sffamily I\normalfont} = 
\int \textrm{d}t \int \textrm{d}^3x ( p \circ \dot{h} - \alpha H - \beta^i H_{i})
\label{VADM} 
$$
$$
 H \equiv \frac{1}{\sqrt{h}}\left(p_{ij} p^{ij} - \frac{p^2}{2}\right) - \sqrt{h}R  = 0,
\label{Vham} \mbox{ } , \mbox{ }
 H_i \equiv -2D_j{p_i}^j = 0 ,
\label{Vmom}
$$
that corresponds to slicing spacetime into a sequence of 3-spaces.  $H$ and $H_i$ are the Hamiltonian 
and momentum constraints.  Their presence indicates redundancy in the $(h_{ij}, p^{ij})$ description of 
GR.  $H_i$ just indicates that the physically relevant information is not in whichever 3-space 
coordinate grid is used, but rather in the shape of the 3-space itself (its 3-geometry).  Thus GR may 
be interpreted as a theory of evolving 3-geometries (`geometrodynamics').  $H$ corresponds to GR 
spacetime being invariant under changes of slicing (corresponding to different choices of time w.r.t 
which this dynamics is understood to occur).  Within any particular slicing this invariance is hidden, 
and we do not know if or how dynamics can be disentangled from the choice of time: one has a 
Problem of Time (POT) in geometrodynamics.  This contributes much to rendering quantum 
geometrodynamics intractable.    

So $H$ is central to geometrodynamics.  Wheeler asked\cite{W} whether its form could be accounted for not 
from rearranging GR but rather from ``plausible first principles".  By allowing embeddability 
into spacetime to dictate the constraint algebra, the form of $H$ was explained\cite{HKT} 
using the algebra of (true and mere grid-stretching) deformations as plausible first 
principles.    

However, recently 3-space -- rather than spacetime -- principles were proposed\cite{BFO}.   
These develop Machian relationalism: physical laws whose validity exends to the universe as 
a whole should depend on relative quantites alone and contain no overall notion of time.  In particle 
mechanics, these respectively involve distances and angles alone being meaningful, and the use of the 
time-label reparametrization-invariant (RI) Jacobi principle (which takes the form 
$
\mbox{\sffamily I\normalfont} = 
\int \textrm{d}t \sqrt{\mbox{potential term} \times \mbox{kinetic term} }$
for the kinetic term homogeneous quadratic in its velocities).  Moreover, an indirect choice 
[`best matching' (BM)] is made for the former: the action is not written in terms of relative 
quantities, but rather auxiliaries associated with absolute (rather than relative) motion are 
adjoined as corrections to the velocities.  Then auxiliary-variation gives constraints which 
renders all absolute motion redundant.  Now, it turns out that GR is indeed of this form.  The 
absolute structure na\"{\i}vely in the 3-space coordinate grids is redundant by the 
shift-corrections to the velocities which variationally-encode $H_i$.  
And upon algebraic elimination of the lapse, GR admits the RI BM action\cite{BSW}:
\be
\mbox{\sffamily I\normalfont}  
= \int \textrm{d}\lambda\int \textrm{d}^3x \sqrt{h}\sqrt{RT},
\label{BSW}
\ee
\be
T = (h^{ik}h^{jl} - h^{ij}h^{kl})(\dot{h}_{ij} - \pounds_{\xi}h_{ij})(\dot{h}_{kl} - \pounds_{\xi}h_{kl}),
\label{T}
\ee
Thus in this sense GR is a perfectly Machian theory.  Note now that $H$ arises `by Pythagoras' as a primary constraint.  

The idea is to consider a large class of BM RI actions in place of (\ref{BSW}).  In each case 
$H_i$ is guaranteed from the inbuilt, entirely spatial shift-variation, and a candidate 
$H^{\prime}$ arises `by Pythagoras'.  Now, applying Dirac's procedure\cite{Dirac} to 
$H$ 
\be
(1 - X)s D_i(N^2D^ip) = 0,
\ee
is required for consistency, where $X$ represents the departure from --1 of the relative 
coefficients in the first factor of (\ref{T}) and the potential $sR + \Lambda$ (const.) 
is in use (other potentials tried are all inconsistent).  The $X = 1$ case is a derivation of 
relativity via {\it obtaining} embeddability into spacetime.  But this arises in conjunction with two 
other cases which {\it do not } have spacetime structure.  First, strong gravities\cite{AS} 
($s = 0$), which generalize the strong-coupled limit of GR applicable near the Big Bang.  
Second, conformal theories\cite{BO} in which maximal ($p = 0$) or constant mean curvature (CMC) 
($\frac{p}{h} =$ spatial constant) slicings are privileged; these include our 
{\it conformal gravity}\cite{BO}), and are closely tied to the GR initial value problem (IVP).  

We next seek inclusion of fundamental matter to test the plausibility of our 3-space 
first principles.  We then discover\cite{BFO} \cite{AB} that via the Dirac procedure GR forces 
simple matter fields to share its null cone, and that electromagnetism and Yang--Mills theories 
arise.  I play down claims about the unique selection of these theories in\cite{AV}, but argue 
that Dirac theory and all the interaction terms of particle physics can be included\cite{AV}.  
This follows from Kucha\v{r}'s spacetime split formalism\cite{K} guaranteeing consistency.  
Although na\"{\i}vely this would usually require spacetime {\it tilt} and {\it derivative-coupling} 
kinematics in addition to BM, I show that the entirely-spatial BM kinematics alone suffices to build 
an action for GR coupled to {\it all} of the above matter fields.

The 3-space -- as opposed to spacetime -- ontology, attempts to restore the centrality in dynamics of 
the configuration space of instants rather than the spacetime arena.  This moves toward 
attempting a na\"{\i}ve Schr\"odinger interpretation (NSR) resolution of the POT.  Alas, 
the geodesic principle idea from the analogy with Jacobi is of no aid\cite{AV}.  Also, spacetime does 
not always emerge in the 3-space approach.  It does not in the conformal theories, which have 
nevertheless a number of orthodox features such as locally-Lorentzian physics and IVP's similar to GR.  
I elect to explain here how these and the quantum-cosmologically-relevant strong gravity 
theories avoid some of GR's pitfalls in a number of POT strategies\cite{POT}.  

In GR, use of ($h_{ij}$, $p^{ij}$) variables and imposing $H$ after quantization gives the 
Wheeler--DeWitt equation for which the Schr\"odinger i.p is indefinite and so not 
probabilistic.  If one proceeds by analogy with Klein--Gordon, one is then floored 
by the indefinite sign of the general $R$-potential.  But for conformal gravity, and a 
range of other conformal and strong gravities the Schr\"{o}dinger i.p itself works, 
whereas for the other strong gravities the potential $\Lambda$ is of definite sign.  

Were one able to produce a mythical canonical transformation in GR to pass to separate embedding 
and dynamical variables (an approach with spacetime ontology connotations), by construction 
one would be equipped with a satisfactory i.p.  In fact the York IVP method furbishes 
such a mythical transformation (corresponding to a CMC internal time), but 
this is unuseble because it involves the implicit solution of the Lichnerowicz PDE  
(conformalization of $H$) which is only tractable numerically.  But for strong gravity 
theories the absence of $R$ renders algebraic the corresponding equation, thus opening up this 
route.  

Were I to show that any of our alternatives explain the very early universe or even all of 
nature, then the corresponding differences above would become directly important.  In the absence 
of this, what I am doing is assessing which features of the GR POT obstructions are 
robust to changes of gravitational theory.  These issues are work in progress.

\end{document}